\documentclass[%
reprint,
 amsmath,amssymb,
 aps,
prl,
floatfix,
]{revtex4-1}

\usepackage{graphicx}%for submission
\usepackage{graphicx}% Include figure files
\usepackage{dcolumn}% Align table columns on decimal point
\usepackage{bm}% bold math
\begin{document}

\preprint{APS/123-QED}

\title{Strategies for reducing the light shift in atomic clocks}

\author{Hidetoshi Katori}
\email{katori@amo.t.u-tokyo.ac.jp}
\affiliation{Department of Applied Physics, Graduate School of Engineering, The University of Tokyo, Bunkyo-ku, Tokyo 113-8656, Japan,}
\affiliation{Innovative Space-Time Project, ERATO, Japan Science and Technology Agency, Bunkyo-ku, Tokyo 113-8656, Japan,}
\affiliation{Quantum Metrology Laboratory, RIKEN, Wako-shi, Saitama 351-0198, Japan.}
 
\author{V. D. Ovsiannikov}
\email{ovd@phys.vsu.ru}
\affiliation{Department of Physics, Voronezh State University, 394006 Voronezh, Russia.}%

\author{S. I. Marmo}
\affiliation{Voronezh State University, 394006 Voronezh, Russia.}%

\author{V. G. Palchikov}
\affiliation{FGUP VNIIFTRI, 141570 Mendeleevo, Moscow Region, Russia}%
\affiliation{National Research Nuclear University ``MEPhI,'' Moscow, Russia.}%

\date{\today}% It is always \today, today, 

\begin{abstract}
Recent progress in optical  lattice clocks requires unprecedented precision in controlling systematic uncertainties  at $10^{-18}$ level. Tuning of nonlinear light shifts is shown to reduce lattice-induced clock shift  for wide range of lattice intensity. Based on theoretical multipolar, nonlinear, anharmonic and  higher-order light shifts, we numerically demonstrate possible strategies for Sr, Yb, and Hg clocks to achieve lattice-induced systematic uncertainty below $1\times 10^{-18}$. 

\begin{description}
%\item[Usage]
%Secondary publications and information retrieval purposes.
\item[PACS numbers]
06.30.Ft, 32.60.+i, 37.10.Jk, 42.62.Eh, 42.62.Fi
\end{description}
\end{abstract}

%06.30.Ft Time and frequency
%32.60.+i	Zeeman and Stark effects
%37.10.Jk	Atoms in optical lattices
%42.62.Eh	Metrological applications; optical frequency synthesizers for precision spectroscopy
%42.62.Fi	Laser spectroscopy

\maketitle
\section{introduction}
Last few years have witnessed significant advances in optical clocks to reach uncertainties of $10^{-18}$ level in ion-based clocks \cite{Cho10}  and optical lattice clocks \cite{Blo14,Ush14}.
Hitherto unexplored accuracy of optical clocks  opens up new possibilities in science and technologies, such as probing new physics via possible variation of fundamental constants \cite{Uza03,Ros08,Hun14}, 
and relativistic geodesy to measure gravitational potential differences \cite{Cho10}. 
Evaluations of perturbations on the clock transitions  are indeed at the heart of these endeavors.

Unperturbed transition frequencies have been accessed by extrapolating perturbations to zero, which is straightforward  
if the correction is  proportional to the perturber. Once the dependence becomes nonlinear, such as the blackbody radiation shift that changes as $T^4$ with  temperature \cite{Ita82},  dedicated experimental \cite{Mid12,She12,Ush14} and theoretical investigations \cite{Saf13} are crucial.
Contrarily, nonlinear response is leveraged to make the clock transition frequency insensitive to perturbations in hyper-Ramsey spectroscopy \cite{Yud10}.

Optical lattice clocks aimed at eliminating light-shift perturbations on the clock transition by operating an optical lattice at the magic frequency \cite{Kat03}, which at first glance should  exempt them from  evaluating lattice-laser intensity. 
However, residual light shifts  arising from hyperpolarizability \cite{Bru06} and multipolar effects \cite{Tai08,Wes11}  manifest as leading systematic uncertainties at low $10^{-17}$.
Coupled with atomic motion in the optical lattice \cite{Tai08,Kat09} whose intensity varies in space to confine atoms, these light shifts show intricate nonlinear response to the lattice intensity \cite{Ovs13}, which makes the corrections highly delicate issue.

In this Letter, we propose strategies to eliminate  light shift perturbations without relying on   a conventional 
zero-extrapolation approach. 
We show that lattice laser frequency and light-polarization-dependent hyperpolarizability effect \cite{Tai06} can be used to tailor intensity dependence of light shift. 
We  define an ``operational magic frequency'' to reduce  light shift to less than $1\times 10^{-18}$ for a sufficiently larger intensity variation than that is  necessary for confining atoms.
Numerical calculations for electric-dipole (E1), magnetic-dipole (M1), and electric-quadrupole (E2) polarizabilities and hyperpolarizabilities are presented for the ${}^1S_0-{}^3P_0$ clock transitions in Sr, Yb, and Hg atoms, which are used to demonstrate the feasibility of the proposed strategies.

\begin{figure}
\includegraphics[width=0.9\linewidth]{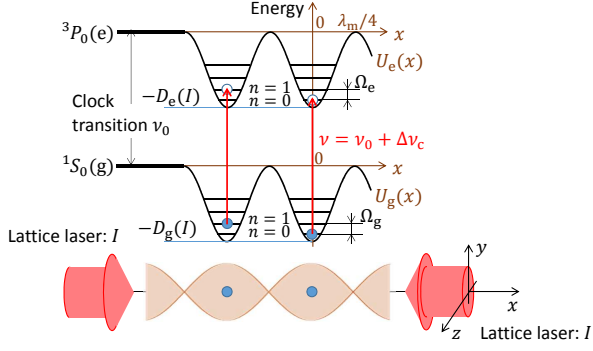}
\caption{\label{fig:schematic} Clock transitions in an optical lattice potential created by counter-propagating laser beams. 
An atom trapped in $U_{\rm g}(x)$ with $|x|\ll\lambda_{\rm m}/4$ is excited on the ${}^1S_0 ({\rm g})\rightarrow {}^3P_0({\rm e})$ clock transition with unperturbed frequency $\nu_0$ and lattice-induced clock  shift $\Delta\nu_{\rm c}$. 
Actual laser beams show transverse intensity distribution $I e^{-2(y^2+z^2)/r_0^2}$ with beam radius $r_0\gg\lambda$, which weakly confines atoms transversely. 
}
\end{figure}

\section{Lattice induced light shifts}
The lattice-induced light shift is given by  the difference between the Stark energies in the ground (g) and excited (e) clock states. 
We assume a one-dimensional optical lattice as depicted in Fig.~\ref{fig:schematic} and consider 
the interaction between a trapped atom and a standing-wave field, 
\begin{equation}
{\bf{E}}(x,t) = 2{{\bf{E}}_0}\cos k x\cos \omega t,
\end{equation}
which  consists of counter-propagating  laser waves with electric field vector ${{\bf{E}}_0}$ (corresponding intensity $I$), frequency $\omega$, and wavevector ${\bf{k}} =\pm k{{\bf{e}}_x}$ with $k = \omega /c$ and $c$ the speed of light.
The atom-lattice interaction is determined by the operator $\hat V(x,t) = {\mathop{\rm Re}\nolimits} [ {\hat V(x)\exp ( - i\omega t)} ]$ with the spatial factor
\begin{equation}
\hat V(x) = {\hat V_{\rm E1}}\cos kx + ({\hat V_{\rm E2}} + {\hat V_{\rm M1}})\sin kx,
\end{equation}
where ${\hat V_{\rm E1}}$, ${\hat V_{\rm E2}}$, and ${\hat V_{\rm M1}}$ correspond to operators of E1, E2, and M1 interactions.

\begin{table}
\caption{\label{tab:table1}
Numerical values of the electric-dipole  ($\alpha_{\rm E1}$),  difference of multipolar  ($\Delta\alpha^{\rm qm}$) and hyper ($\Delta\beta$) polarizabilities for Sr, Yb and Hg clock transitions at the magic wavelengths $\lambda_{\rm m}$. Frequency shifts and vibrational frequencies $\Omega(I)\approx \frac{2}{\hbar}\sqrt {{{\cal E}_{\rm{R}}}\alpha _{{\rm{g}}({\rm{e}})}^{{\rm{dqm}}}I} $ are calculated for an atom trapped at the anti-node of the standing wave, where  $I\, {\rm (kW/cm^2)}$ denotes the intensity of each traveling wave laser.
Non-zero imaginary part of the differential hyperpolarizability $\Delta\beta^{l(c)}$ for Hg accounts for two photon ionization rate in the optical lattice, where $l$ and $c$ correspond to linear and circular lattice-light polarization.
${\cal E}_{\rm R}$ is the lattice photon recoil energy.
A merit factor $\kappa$ indicates the insensitivity to atomic-motion-dependent lattice light shift.
Operational lattice intensity $I_{\rm op}$ assumes to produce 5 times deeper lattice potential depth than the energy of laser-cooled atoms.  
}
\begin{ruledtabular}
\begin{tabular}{cccc}
Atom&Sr&Yb&Hg\\
\hline
$\lambda_{\rm m}\, {\rm (nm)}$&813.4&	759.4&	362.6\\
$\nu_0\,({\rm THz})$&429&		518&	1129\\
${\alpha^{\rm E1}}/h {\left( {\frac{{{\rm{kHz}}}}{{{\rm{kW/c}}{{\rm{m}}^{\rm{2}}}}}} \right)}$&45.2&40.5&5.70\\

${\Delta \alpha^{\rm qm}}/h {\left( {\frac{{{\rm{mHz}}}}{{{\rm{kW/c}}{{\rm{m}}^{\rm{2}}}}}} \right)}$&1.38&$-1.71$&8.25\\
${  {\Delta \beta^l}/h \left( {\frac{{{\rm{\mu Hz}}}}{{{{\left( {{\rm{kW/c}}{{\rm{m}}^{\rm{2}}}} \right)}^2}}}} \right)}$&$-200$&$-309$&$-2.20+0.82 i$\\
${ {\Delta \beta^c}/h \left( {\frac{{{\rm{\mu Hz}}}}{{{{\left( {{\rm{kW/c}}{{\rm{m}}^{\rm{2}}}} \right)}^2}}}} \right)}$&$-311$&238&$4.40+1.21 i$\\
${\frac{{{\Omega}}}{{2\pi \sqrt {{I}} }}\left( {\frac{{{\rm{kHz}}}}{{\sqrt {{\rm{kW/c}}{{\rm{m}}^{\rm{2}}}} }}} \right)}$&25.05&18.03&13.1\\
$1\cdot10^9\cdot{\frac{{\partial  {\Delta \alpha^{\rm E1}} }}{{ h \partial \nu }}\left( \frac{1}{\rm{kW/cm}^2} \right)}$& 0.254&0.720&0.134\\
${\cal E}_{\rm R}/h\, {\rm{ (kHz)}}$&3.47&		2.00&	7.57\\
$\kappa= \alpha^{\rm E1}/|\Delta \alpha^{\rm qm}|$ & $3.3\times10^7$ & $2.4\times10^7$ & $6.9\times10^5$ \\
$I_{\rm op}= 5k_{\rm B} T/\alpha^{\rm E1}\,({\rm kW/cm^2}) $ & 2.3 & 10 & 550 
\end{tabular}
\end{ruledtabular}
\end{table}

The second- and fourth-order terms of atom-lattice interaction energy $\hat V(x)$ correspond to linear and quadratic terms in  lattice-laser intensity $I$.
The optical lattice potential for an atom at $|x|\ll \lambda=2\pi/k$ (see Fig.~\ref{fig:schematic}) is given by \cite{Ovs13}
\begin{equation}
U_{\rm g(e)}(x,I) \approx  - {D_{\rm g(e)}}(I) + u_{\rm g(e)}^{(2)}(I){x^2} - u_{\rm g(e)}^{(4)}(I){x^4} +...
\label{eq:potential}
\end{equation}
with potential depth  given by
\begin{equation}
{D_{\rm g(e)}}(I) =-U_{\rm g(e)}(0,I) =  \alpha _{\rm g(e)}^{\rm E1}(\omega )I + \beta _{\rm g(e)}^{}(\omega,\xi ){I^2},
\label{eq:depth}
\end{equation}
which is determined by the electric-dipole polarizability $\alpha _{{\rm{g(e)}}}^{{\rm{E1}}}(\omega )$ and hyperpolarizability $\beta_{\rm g(e)}(\omega,\xi)$ with $\xi$ the degree of circular polarization of light as discussed later.
The coefficient for the harmonic term in Eq.~(\ref{eq:potential}), 
$u_{\rm g(e)}^{(2)}(I) 
= {\textstyle{1 \over 2}}M\Omega _{{\rm{g}}({\rm{e}})}^2(I)$, 
determines the vibrational frequency ${\Omega _{\rm g(e)}}(I)$ of atoms in the lattice. 
In terms of the photon recoil energy ${{\cal E}_{\rm{R}}} = {(\hbar k)^2}/2{\cal M}$ with ${\cal M}$ the atomic mass and $\hbar=h/2\pi$ the Planck constant, the vibrational frequency is  given by  \cite{Ovs13}
\begin{equation}
{\Omega _{{\rm{g(e)}}}}(I) = \frac{2}{\hbar}\sqrt {{{\cal E}_{\rm{R}}}\left[ {\alpha _{{\rm{g(e)}}}^{{\rm{dqm}}}(\omega )I + 2\beta _{{\rm{g(e)}}}^{}(\omega,\xi  ){I^2}} \right]},
\label{eq:HO}
\end{equation}
where a combined E1-E2-M1 polarizability
\begin{equation}
\alpha _{{\rm {g(e)}}}^{{\rm dqm}}(\omega ) = \alpha _{{\rm {g(e)}}}^{{\rm{E1}}}(\omega ) - \alpha _{{\rm{g(e)}}}^{{\rm{qm}}}(\omega)
\end{equation}
is the difference between E1 polarizability and the sum of E2 and M1 polarizabilities 
$\alpha _{{\rm{g}}({\rm{e}})}^{{\rm{qm}}}(\omega ) = \alpha _{{\rm{g}}({\rm{e}})}^{{\rm{E2}}}(\omega ) + \alpha _{{\rm{g}}({\rm{e}})}^{{\rm{M1}}}(\omega )$. 
The lowest-order anharmonic correction in the lattice  is given by 
\begin{equation}
u_{\rm g(e)}^{(4)}(I) = \left[ {\alpha _{\rm g(e)}^{\rm dqm}(\omega )I + 5\beta _{\rm g(e)}^{}(\omega,\xi ){I^2}} \right]\frac{{{k^4}}}{3}.
\end{equation}

The energy of an atom in the $n$-th vibrational state $|n\rangle$ is calculated to be 
\begin{multline}
{E}_{{\rm{g(e)}}}^{{\rm{vib}}}(I,n) 
= - D_{{\rm{g(e)}}}^{}(I)
\\ + \hbar {\Omega _{{\rm{g(e)}}}}(I)\left( {n + {\textstyle{1 \over 2}}} \right) { - E}_{{\rm{g(e)}}}^{(4)}(I)\left( {{n^2} + n + {\textstyle{1 \over 2}}} \right),
\label{eq:energy}
\end{multline}
where the second term corresponds to the harmonic-oscillator energy and the last term the anharmonic correction given by 
\begin{equation} 
{E}_{{\rm{g(e)}}}^{{\rm{(4)}}}(I) = \frac{{{{\cal E}_{\rm{R}}}}}{2}\left[ {1 + \frac{{3{\beta _{{\rm{g(e)}}}}(\omega,\xi )I}}{{\alpha _{{\rm{g(e)}}}^{{\rm{dqm}}}(\omega )}}} \right].
\label{eq:anharmonic}
\end{equation}
The lattice-induced clock light shift is given by the difference of  energies [Eq.~(\ref{eq:energy})] for the atom in its ground and excited states with vibrational state $|n\rangle$ being unchanged (the Lamb-Dicke regime), 
%The clock shift is given by
\begin{multline}
h\Delta \nu _{\rm{c}}(I,n) = {E}_{\rm{e}}^{{\rm{vib}}}(I,n) - {E}_{\rm{g}}^{{\rm{vib}}}(I,n)
 =  - \Delta D(I) \\
+ \hbar \Delta \Omega (I)\left( {n + {\textstyle{1 \over 2}}} \right)
-  \Delta {{E}^{(4)}}(I)\left( {{n^2} + n + {\textstyle{1 \over 2}}} \right),
\label{eq:clock_shift}
\end{multline}
where we define $\Delta D(I) = {D_{\rm e}}(I) - {D_{\rm g}}(I)$, $\Delta \Omega (I) = {\Omega _{\rm e}}(I) - {\Omega _{\rm g}}(I)$, and $\Delta {{E}^{(4)}}(I) = {E}_{\rm e}^{(4)}(I) - {E}_{\rm g}^{(4)}(I)$.

The hyperpolarizabilities depend on the lattice-laser ellipticity in addition to  its frequency, and are given by
\begin{equation}
{\beta _{{\rm{g(e)}}}}(\omega ,\xi ) = \beta _{{\rm{g(e)}}}^{{l}}(\omega) + {\xi ^2}\left[ {\beta _{{\rm{g(e)}}}^{{c}}(\omega) - \beta _{{\rm{g(e)}}}^{{l}}(\omega)}\right],
\label{eq:hyp_pol}
\end{equation}
where $\beta _{{\rm{g(e)}}}^{{l(c)}}(\omega )$  is the hyperpolarizability for linear (circular) polarized light.
The degree of  circular polarization is defined by $\xi=\sin {2\chi}$, where $\tan\chi$ determines  the ratio of the minor to  major  axis of the polarization ellipse with ellipticity angle defined in $0\leq\chi\leq\pi/4$.
When  $\Delta\beta^l(=\beta_{\rm e}^l-\beta_{\rm g}^l)$ and $\Delta\beta^c(=\beta_{\rm e}^c-\beta_{\rm g}^c)$ have opposite signs, there exists a ``magic ellipticity'' determined by ${\xi _{\rm m}} =  1/\sqrt {1 - \Delta {\beta ^c}/\Delta {\beta ^l}} $   \cite{Tai06}, which eliminates  the differential hyperpolarizability $\Delta \beta (\xi ) = \Delta {\beta ^l} + {\xi ^2}(\Delta {\beta ^c} - \Delta {\beta ^l})$  [Eq.~(\ref{eq:hyp_pol})]. 
 However, a more important consequence for the following discussion is the tunability of 
$\Delta \beta(\xi)$ between $\Delta {\beta ^l}$ and $\Delta {\beta ^c}$.

\begin{figure}[]
\includegraphics[width=\linewidth]{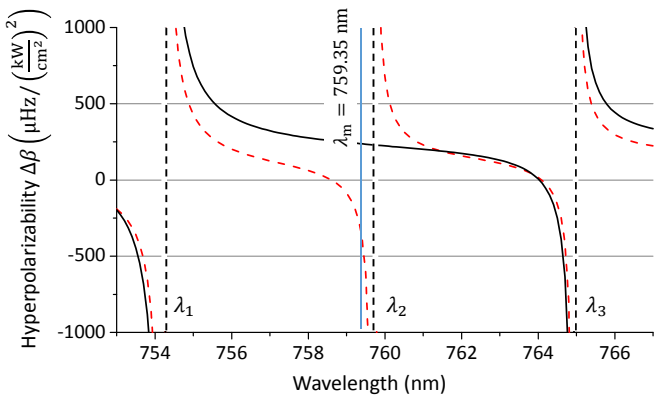}
\caption{\label{fig:hyp_Yb} The wavelength-dependent hyperpolarizability for Yb  for  linear ($\Delta \beta^l$, dashed) and circular ($\Delta \beta^c$, solid) polarized lattice near the magic wavelength $\lambda_{\rm m}$. The vertical dashed-lines indicate two-photon resonances. 
}
\end{figure}

Figure~\ref{fig:hyp_Yb} shows  the  hyperpolarizability $\Delta \beta^{l(c)}$ for Yb atoms near the magic wavelength. The two-photon resonance on the $6s6p\,{}^3P_0\rightarrow 6s8p\,{}^3P_0$ transition at $\lambda_2=759.71\,{\rm nm}$ appears only for linear polarized lattice, which locates between $\lambda_1=754.23$~nm and $\lambda_3=764.95$~nm that correspond to two-photon resonances on the  $6s6p\,{}^3P_0\rightarrow 6s8p\,{}^3P_2,\, 6s5f\,{}^3F_2$ transitions \cite{Bar08}. 
These resonances allow $\Delta \beta^l$  and $\Delta \beta^c$  to have  opposite signs. Similar discussion holds for Hg, however, not for Sr.
Table~\ref{tab:table1} summarizes  the susceptibilities for Sr, Yb, and Hg clock transitions calculated in the model-potential approximation \cite{Man86}. 
While some of these values require experimental investigations, in the following, we apply them to demonstrate the concept of the operational magic wavelength.

\section{Operational magic frequency}

In order to clarify the light-shift dependence on intensity $I$, we approximate  the light shift [Eq.~(\ref{eq:clock_shift})] assuming  experimentally feasible lattice laser intensities to trap laser-cooled  atoms as listed in Table~\ref{tab:table1}.
Following quantities, (i) the E2-M1  polarizabilities $\alpha_{\rm g(e)}^{\rm qm}=\alpha_{\rm g(e)}^{\rm E2}+\alpha_{\rm g(e)}^{\rm M1}$,  (ii) the hyperpolarizability effect $\beta_{\rm g(e)} I$, 
and (iii) the differential dipole polarizability $\Delta\alpha^{\rm E1}=\alpha_{\rm e}^{\rm E1}-\alpha_{\rm g}^{\rm E1}=\frac{{\partial \Delta {\alpha ^{\rm E1}}}}{{\partial \omega }}\delta \omega$,   are about  $10^{6}$ times smaller than the electric-dipole polarizability $\alpha_{\rm g(e)}^{\rm E1}\approx \alpha^{\rm E1}$. The light  shift is then  expanded in Taylor series in the vicinity of $\omega\approx \omega_{\rm m}^{\rm E1}$ and neglecting higher order terms, 
\begin{multline}
h\Delta {\nu _{\rm{c}}}(I,n,\delta \nu ,\xi ) = {c_{1/2}}{I^{1/2}} + {c_1}I + {c_{3/2}}{I^{3/2}} + {c_2}{I^2}\\
 \approx \left( {\frac{{\partial \Delta {\alpha ^{{\rm{E1}}}}}}{{\partial \nu }}\delta \nu  - \Delta {\alpha ^{{\rm{qm}}}}} \right)\left( {2n + 1} \right)\sqrt {\frac{{{{\cal E}_{\rm{R}}}}}{{4{\alpha ^{{\rm{E1}}}}}}} {I^{1/2}} \\
- \left[ {\frac{{\partial \Delta {\alpha ^{{\rm{E1}}}}}}{{\partial \nu }}\delta \nu  + \Delta \beta (\xi )\left( {2{n^2} + 2n + 1} \right)\frac{{3{{\cal E}_{\rm{R}}}}}{{4{\alpha ^{{\rm{E1}}}}}}} \right]I \\
+ \Delta \beta (\xi )\left( {2n + 1} \right)\sqrt {\frac{{{{\cal E}_{\rm{R}}}}}{{{\alpha ^{{\rm{E1}}}}}}} {I^{3/2}} - \Delta \beta (\xi ){I^2},
\label{eq:poly}
\end{multline}
where $\delta \nu(=\delta \omega/2\pi)$ is  detuning from the ``E1-magic frequency'' defined by $\Delta\alpha^{\rm E1}(\omega_{\rm m}^{\rm E1})=0$, $\Delta \alpha^{\rm qm}=\alpha_{\rm e}^{\rm qm}-\alpha_{\rm g}^{\rm qm}$ is the differential multipolar polarizability.

The magic frequency $\omega_{\rm m}/2\pi(=c/\lambda_{\rm m})$ so far aimed at  minimizing $c_1$, which dominates Eq.~(\ref{eq:poly}), by tuning $\omega_{\rm m}\rightarrow \omega_{\rm m}^{\rm E1}$ \cite{Kat03}.
However, this protocol is no longer valid for fractional uncertainty  $\Delta \nu_{\rm c}/\nu_0\sim 10^{-17}$ as the other $c_{j} I^{j}$ terms equally contribute \cite{Tai08}. 
We may define a merit factor $\kappa\equiv\alpha^{\rm E1}/|\Delta \alpha^{\rm qm}|$ in Table~\ref{tab:table1} to indicate applicability of the ``E1-magic frequency'' by neglecting multipolar effect.
$\kappa^{-1}$ indicates the fractional contribution of the E2-M1 effect that introduces atomic-motion induced $I^{1/2}$ nonlinearity. 
Since typical clock experiments \cite{Blo14,Ush14} are  performed in a relatively low intensity regime ($\Delta \beta I\leq\Delta\alpha^{\rm qm}$), a large $\kappa$ for Sr and Yb may validate  a linear extrapolation of the clock shifts down to $10^{-17}$ level. 
However, as the merit factor decreases for Hg, this simple approach breaks down and new strategies are required. 

\begin{figure}[]
\includegraphics[width=\linewidth]{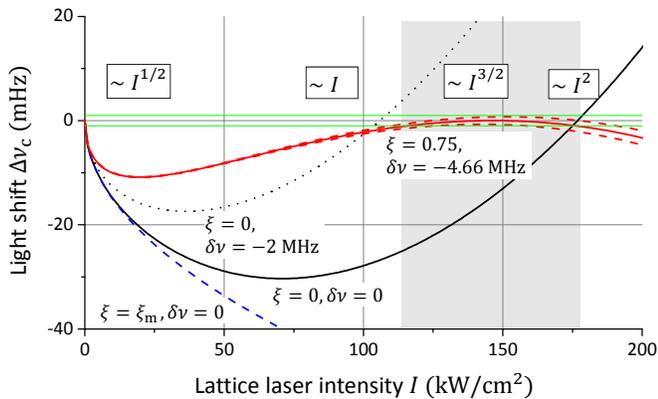}
\caption{\label{fig:Hg} Light shift for a Hg clock 
as a function of  laser intensity $I$. Black-solid and blue-dashed lines correspond to linear-polarized   and ``magic-elliptical'' light with detuning $\delta \nu=0$. With $\delta \nu= -4.66\, {\rm MHz}$ and $\xi^{\rm Hg}=0.75$, the  light shift (red-solid) becomes less than $\pm1$~mHz (green) for  gray shaded region. Red-dashed lines indicate  tolerance (0.5\%) for $\xi^{\rm Hg}$. For linear polarized light ($\xi=0$) with  $\delta\nu=-2\,{\rm MHz}$, the light shift becomes insensitive to $\Delta I$ around $I\sim36\,{\rm kW/cm^2}$ (dotted line).  }
\end{figure}

Figure~\ref{fig:Hg} illustrates  intensity dependence of the light shift $\Delta \nu_{\rm c}^{\rm Hg}(I)$ for Hg. 
For low laser intensity, $I^{1/2}$ behavior dominates, whose coefficient $c_{1/2}$ is determined by the  electric-dipole $(\frac{{\partial \Delta {\alpha ^{{\rm{E1}}}}}}{{\partial \nu }}\delta \nu )$ and multipolar $( \Delta \alpha ^{\rm{qm}})$ polarizabilities. 
For intermediate intensity where clocks  operate, the leading  term $\propto I$  is  determined by the electric-dipole polarizability and slightly by the hyperpolarizability $\Delta \beta(\xi)$ effect via the anharmonic correction [Eq.~(\ref{eq:anharmonic})], in addition, the $I^{3/2}$ and $I^2$ terms depend on $\Delta \beta(\xi)$.

\begin{figure}[]
\includegraphics[width=\linewidth]{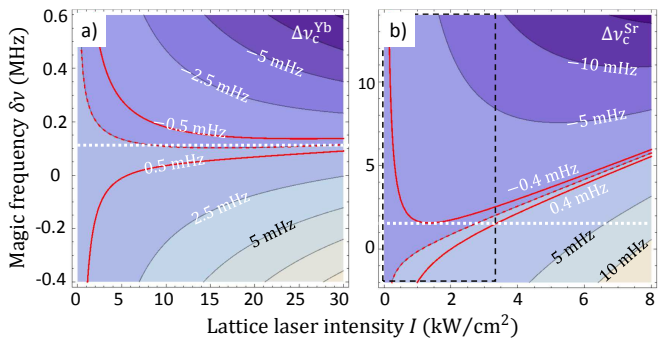}
\caption{\label{fig:Yb} Contour plots of  light shifts  for a) Yb and b) Sr clock transitions as functions of lattice laser intensity $I$ and  detuning $\delta \nu$, for $\xi^{\rm Yb}=0.75$ and $\xi^{\rm Sr}=0$. The red-dotted lines show zero light shift.
The region bound by red lines corresponds to light shift $|\Delta \nu_{\rm c}|/\nu_0\leq1\times 10^{-18}$.
Tuning to the operational magic frequency  $\delta \nu$ as indicated by white dotted lines, wide range of operational intensities are allowed.
} 
\end{figure}

It is apparent that there is no ``magic condition''  that  sets four coefficients $c_j\rightarrow0$, as there are  only three free parameters $n$, $\delta \nu$ and $\xi$,  available.
Consider that an operational lattice intensity $I_{\rm op}$ is naturally determined (Table~\ref{tab:table1}) so as to trap  Doppler-cooled atoms  $D_{\rm g(e)}\approx \alpha^{\rm E1}I_{\rm op}\sim 5\, k_{\rm B} T$  with temperatures of $T\sim1\,{\rm \mu}$K, $4\,{\rm \mu}$K, and $30\,{\rm \mu}$K for Sr, Yb, and Hg, respectively \cite{Muk03,Lem09, Hac08}.
We may then tune  $\delta \nu$ and  $\xi$  to minimize the  light shift $\Delta \nu_{\rm c}$ for $I_{\rm op}$ with its allowance $\Delta I$ as wide as possible, as the laser intensity is spatially non-uniform. 
Actual one-dimensional optical lattices employ Gaussian beams with an intensity profile  $I e^{-2(y^2+z^2)/r_0^2}$ and a beam radius $r_0(\gg\lambda)$ to confine thermal motion of atoms in the transverse direction (see  Fig.~\ref{fig:schematic}), which inevitably introduces intensity inhomogeneity of $\Delta I/I\approx k_{\rm B}T/D_{\rm e(g)}$.  

In the following, to simplify the discussion, we assume atoms in the vibrational ground state $(n=0)$ along $x$-axis by applying a sideband cooling. 
We then  look for  conditions that satisfy  both  ${\left. {\frac{{\partial \Delta {\nu _{\rm c}}(I,\delta \nu, \xi) }}{{\partial {I}}}} \right|_{{I} = {I_{\rm op}}}} = 0$ and $\Delta\nu_{\rm c}(I_{\rm op},\delta \nu, \xi)=0$.
The  red solid line in Fig.~\ref{fig:Hg} shows the light shift $\Delta \nu_{\rm c}^{\rm Hg}(I)$ 
with $\delta \nu=-4.66\,{\rm  MHz}$ and $\xi^{\rm Hg}=0.75$, which demonstrates that the light shift becomes less than 1~mHz (corresponding to fractional uncertainty of $1\times 10^{-18}$) for  
 $115\, {\rm kW/cm^2}<I<177\, {\rm kW/cm^2}$. Here, the two photon ionization rate  ${\mathop{\rm Im}\nolimits} [ \beta /h] I_{{\rm{op}}}^2\sim0.02\,{\rm Hz}$ for $I_{\rm op}\sim 150\, {\rm kW/cm^2}$ is negligible for obtaining a few Hz linewidth.

Figure~\ref{fig:Yb}a calculates the light shift $\Delta \nu_{\rm c}^{\rm Yb}(I,\delta \nu)$ of Yb as functions of lattice laser intensity $I$ and detuning $\delta \nu$ with $\xi^{\rm Yb}=0.75$, where a slight difference from the magic ellipticity $\xi_{\rm m}^{\rm Yb}=0.7516$ compensates the multipolar effect. 
Taking a lattice-laser detuning of $\delta \nu=0.11\,{\rm MHz}$ (white dotted line), light shift uncertainty $|\Delta \nu_{\rm c}^{\rm Yb}(I,\delta \nu)|/\nu_0^{\rm Yb}$ can be well less than $1\times 10^{-18}$ for an entire lattice laser intensity $I$ shown in the plot.  
For 0.5\% variation in  $\xi^{\rm Yb}$, corresponding intensity range is reduced to $ I<12\,{\rm kW/cm^2}$ by allowing fractional uncertainty to $1\times 10^{-18}$.
The vector light shift arising from the elliptical light may be canceled out by averaging the clock transitions with Zeeman substates $m=\pm1/2$ \cite{Tak06}. 
$^{171}$Yb and $^{199}$Hg  with nuclear spin of $ 1/2$ are the best candidate for this scheme because of the lack of tensor light shift.

The sign of hyperpolarizability is not tunable for Sr. However, as shown by the region bound by a dashed rectangle in Fig.~4b, $|\Delta \nu _{\rm{c}}^{{\rm{Sr}}}|/\nu _0^{{\rm{Sr}}} < 1 \times {10^{ - 18}}$ holds for $0<I<3.3\, {\rm kW/cm^2}$ for a linear-polarized lattice  ($\xi=0$) with $\delta\nu=1.5\, {\rm MHz}$, which makes Sr  an attractive candidate. 
Its very low Doppler temperature allows  Sr lattice  to operate at low intensity $I\approx 2.3\, {\rm kW/cm^2}$, where high merit factor $\kappa_{\rm Sr}$ keeps the multipolar effect  small and the hyperpolarizability effect does not come into play. 
Similar low intensity optimization is applicable to Yb and Hg at the expense of lattice-trapped atoms or by applying deep laser-cooling  on the $^1S_0-{}^3P_0$ clock transitions.

In case the hyperpolarizability is not tunable and/or experimental  issues require to use linear-polarized lattice ($\xi=0$), an optimal detuning $\delta\nu_{\rm op}$ may be determined as ${\left. {\frac{{\partial \Delta {\nu _{\rm{c}}}(I,\delta \nu_{\rm op} )}}{{\partial I}}} \right|_{I = {I_{{\rm{op}}}}}} = 0$. Although the clock transition may  suffer from finite correction  $\Delta\nu_{\rm c}(I_{\rm op},\delta\nu_{\rm op})$, the scheme would also work as the light shift becomes insensitive to lattice intensity $\Delta I/I$ for a certain range as seen in black-solid and dotted lines in Fig.~\ref{fig:Hg}. 

\section{Summary and outlook}

We have proposed  an operational magic frequency  that makes   light shift smaller than  $1\times 10^{-18}$ for a wide range of  lattice  intensity as a result of cancellations of light shifts of different origin. 
Numerically demonstrated   operational intensities with $\Delta I/I>40\%$ offer a robust protocol to guarantee  superb reproducibility of optical lattice clocks.
It is noticeable that this intensity allowance can be larger than the intensity variation arising from the thermal motion of atoms in the transverse direction  $\Delta I/I\approx k_{\rm B}T/D_{\rm e(g)}$, as it can be  in the range of $0.1-0.2$ for atoms in  thermal equilibrium \cite{Oha01}.
Moreover all the tuning parameters, both the lattice-laser intensity $I$ and the degree of circular polarization $\xi$, can be spectroscopically identified   via the vibrational frequency $\Omega$ and the vector light shift, respectively, allowing one to define $I$ and $\xi$ in terms of frequency.
In order to apply the proposed scheme, experimental investigations of the polarizabilities are crucial, where 
recently demonstrated clock reproducibility at $2\times 10^{-18}$ \cite{Ush14} can be a powerful tool. 
With  precise determination of lattice intensity  $\Delta I/I$, e.g., by further cooling the transverse motion of atoms down to the photon-recoil temperatures \cite{Muk03},   clock uncertainties of low  $10^{-19}$ will be in scope.

\section{Acknowledgments}
We acknowledge M. Takamoto, K. Yamanaka, N. Ohmae, M. Das and N. Nemitz for useful comments.
VDO acknowledges financial support from the RF Ministry of Education and Science (Project No.1226) and from RFBR (Grant No.14-02-00516-a).

\bibliography{OP_MWL}% Produces the bibliography via BibTeX.

\end{document}